\documentclass[iop,twocolumn]{emulateapj}
\usepackage[T1]{fontenc}
\usepackage{microtype}
\usepackage{afterpage,amsmath,amssymb,graphicx,natbib}
\usepackage[colorlinks,urlcolor=blue,citecolor=blue,linkcolor=blue]{hyperref}
\newcommand\plotonebig[2]{\centering \leavevmode
\includegraphics[width=#2\linewidth]{#1}}

\newcommand\apx{\ensuremath{\sim}}
\newcommand\chandra{\textit{Chandra}}
\newcommand\citeeg[1]{\citep[e.g.,][]{#1}}

\newcommand\gsim{\gtrsim}
\newcommand\ha{{\ensuremath{\text{H}\alpha}}}
\newcommand\kms{km~s$^{-1}$}
\newcommand\Lb{\ensuremath{L_\text{bol}}}
\newcommand\LhLb{\ensuremath{L_\ha/\Lb}}
\newcommand\Lx{\ensuremath{L_\text{X}}}
\newcommand\LxLb{\ensuremath{L_\text{X}/\Lb}}
\newcommand\lsim{\lesssim}

\newcommand\mbv{\ensuremath{\langle B_V\rangle}}
\newcommand\Msun{\ensuremath{M_\odot}}

\newcommand\prot{\ensuremath{P_\text{rot}}}
\newcommand\ro{\ensuremath{\text{Ro}}}
\newcommand\rco{\ensuremath{R_\text{corot}}}

\newcommand\tauc{\ensuremath{\tau_c}}
\newcommand\teff{\ensuremath{T_\text{eff}}}
\newcommand\vsi{\ensuremath{v \sin i}}

\newcommand\simpfx{http://simbad.u-strasbg.fr/simbad/sim-id?Ident=}
\newcommand{\MakeObj}[5]{% {ident}{url-escaped}{longname}{shortname}{truncname}
  \expandafter\newcommand\csname pkgwobjl#1\endcsname{\href{\simpfx #2}{#3}}%
  \expandafter\newcommand\csname pkgwobjs#1\endcsname{\href{\simpfx #2}{#4}}%
  \expandafter\newcommand\csname pkgwobjn#1\endcsname{\href{\simpfx #2}{#4}}%
  \expandafter\newcommand\csname pkgwobjt#1\endcsname{\href{\simpfx #2}{#5}}}
\newcommand{\objl}[1]{\csname pkgwobjl#1\endcsname}
\newcommand{\obj}[1]{\csname pkgwobjs#1\endcsname}
\newcommand{\objn}[1]{\csname pkgwobjn#1\endcsname}
\newcommand{\objt}[1]{\csname pkgwobjt#1\endcsname}

\defcitealias{paper1}{Paper~\textsc{i}}
\newcommand\paperone{\citetalias{paper1}}

\MakeObj{1rxsj115928.5-524717}{1RXS\%20J115928.5-524717}{1RXS\,J115928.5-524717}{1RXS\,J115928.5-524717}{1RXS\,J115928.5-524717}
\MakeObj{2massij0523382-140302}{2MASSI\%20J0523382-140302}{2MASSI\,J0523382-140302}{2MASSI\,J0523382-140302}{2MASSI\,J0523382-140302}
\MakeObj{2massj07464256+2000321ab}{2MASS\%20J07464256\%2b2000321\%20}{2MASS\,J07464256+2000321\,AB}{2MASS\,J07464256+2000321\,AB}{2MASS\,J07464256+2000321\,AB}
\MakeObj{2massj10481463-3956062}{2MASS\%20J10481463-3956062}{2MASS\,J10481463-3956062}{2MASS\,J10481463-3956062}{2MASS\,J10481463-3956062}
\MakeObj{2masswj1507476-162738}{2MASSW\%20J1507476-162738}{2MASSW\,J1507476-162738}{2MASSW\,J1507476-162738}{2MASSW\,J1507476-162738}
\MakeObj{2mucd}{2MUCD\%2010158}{2MUCD\,10158}{2MUCD\,10158}{2MUCD\,10158}
\MakeObj{brib0021-0214}{BRI\%20B0021-0214}{BRI\,B0021-0214}{BRI\,B0021-0214}{BRI\,B0021-0214}
\MakeObj{dp0255}{DENIS-P\%20J025503.3-470049}{DENIS-P\,J025503.3$-$470049}{DENIS 0255$-$4700}{xx}
\MakeObj{denis-pj1228.2-1547ab}{DENIS-P\%20J1228.2-1547}{DENIS-P\,J1228.2-1547\,AB}{DENIS-P\,J1228.2-1547\,AB}{DENIS-P\,J1228.2-1547\,AB}
\MakeObj{g208-44}{G\%20208-44}{G\,208--44\,AB}{G\,208--44\,AB}{G\,208--44}
\MakeObj{g208-45}{G\%20208-45}{G\,208--45}{G\,208--45}{??}
\MakeObj{gj1245ac}{GJ\%201245A}{GJ\,1245\,AC}{GJ\,1245\,AC}{AC}
\MakeObj{gj1245a}{GJ\%201245A}{GJ\,1245\,A}{GJ\,1245\,A}{A}
\MakeObj{gj1245b}{GJ\%201245B}{GJ\,1245\,B}{GJ\,1245\,B}{B}
\MakeObj{gj1245c}{GJ\%201245C}{GJ\,1245\,C}{GJ\,1245\,C}{C}
\MakeObj{gj1245}{GJ\%201245}{GJ\,1245}{GJ\,1245}{GJ\1245}
\MakeObj{gl-166c}{Gl-166\%20C}{Gl-166\,C}{Gl-166\,C}{Gl-166\,C}
\MakeObj{gl206}{Gl\%20206}{Gl\,206}{Gl\,206}{Gl\,206}
\MakeObj{gl388ab}{Gl\%20388\%20}{Gl\,388\,AB}{Gl\,388\,AB}{Gl\,388\,AB}
\MakeObj{gl406}{Gl\%20406}{Gl\,406}{Gl\,406}{Gl\,406}
\MakeObj{gl494ab}{Gl\%20494\%20}{Gl\,494\,AB}{Gl\,494\,AB}{Gl\,494\,AB}
\MakeObj{gl569ba}{Gl\%20569B}{Gl\,569Ba}{Gl\,569Ba}{Gl\,569Ba}
\MakeObj{gl65-ab}{Gl\%2065-}{Gl\,65-AB}{Gl\,65-AB}{Gl\,65-AB}
\MakeObj{gl669b}{Gl\%20669b}{Gl\,669b}{Gl\,669b}{Gl\,669b}
\MakeObj{gl729}{Gl\%20729}{Gl\,729}{Gl\,729}{Gl\,729}
\MakeObj{gl735}{Gl\%20735}{Gl\,735}{Gl\,735}{Gl\,735}
\MakeObj{gl860b}{Gl\%20860b}{Gl\,860b}{Gl\,860b}{Gl\,860b}
\MakeObj{gl896b}{Gl\%20896b}{Gl\,896b}{Gl\,896b}{Gl\,896b}
\MakeObj{hd130948b}{HD\%20130948b}{HD\,130948b}{HD\,130948b}{HD\,130948b}
\MakeObj{hd130948c}{HD\%20130948c}{HD\,130948c}{HD\,130948c}{HD\,130948c}
\MakeObj{kelu-1ab}{Kelu-1\%20}{Kelu-1\,AB}{Kelu-1\,AB}{Kelu-1\,AB}
\MakeObj{lhs248}{LHS\%20248}{LHS\,248}{LHS\,248}{LHS\,248}
\MakeObj{lhs292}{LHS\%20292}{LHS\,292}{LHS\,292}{LHS\,292}
\MakeObj{lhs523}{LHS\%20523}{LHS\,523}{LHS\,523}{LHS\,523}
\MakeObj{lhs2065}{LHS\%202065}{LHS\,2065}{LHS\,2065}{LHS\,2065}
\MakeObj{lhs2397}{LHS\%202397a}{LHS\,2397a\,AB}{LHS\,2397a\,AB}{LHS\,2397a}
\MakeObj{lhs3003}{LHS\%203003}{LHS\,3003}{LHS\,3003}{LHS\,3003}
\MakeObj{lhs3406}{LHS\%203406}{LHS\,3406}{LHS\,3406}{LHS\,3406}
\MakeObj{lp349}{LP\%20349-25}{LP\,349--25\,AB}{LP\,349--25\,AB}{LP\,349--25}
\MakeObj{lp412}{LP\%20412\%2031}{LP\,412\,31}{LP\,412\,31}{LP\,412\,31}
\MakeObj{lp647}{LP\%20647-13}{LP\,647--13}{LP\,647--13}{LP\,647--13}
\MakeObj{lp851}{LP\%20851-346}{LP\,851--346}{LP\,851--346}{LP\,851--346}
\MakeObj{lp944}{LP\%20944-20}{LP\,944--20}{LP\,944--20}{LP\,944--20}
\MakeObj{lspmj0036+1821}{LSPM\%20J0036\%2b1821}{LSPM\,J0036+1821}{LSPM\,J0036+1821}{LSPM\,J0036+1821}
\MakeObj{lspmj1835+3259}{LSPM\%20J1835\%2b3259}{LSPM\,J1835+3259}{LSPM\,J1835+3259}{LSPM\,J1835+3259}
\MakeObj{lsrj0602+3910}{LSR\%20J0602\%2b3910}{LSR\,J0602+3910}{LSR\,J0602+3910}{LSR\,J0602+3910}
\MakeObj{n33370}{NLTT\%2033370\%20}{NLTT\,33370\,AB}{NLTT\,33370\,AB}{NLTT\,33370\,AB}
\MakeObj{n40026}{NLTT\%2040026}{NLTT\,40026}{NLTT\,40026}{NLTT\,40026}
\MakeObj{pc0025+0447}{2MASS\%20J00274197\%2b0503417}{PC\,0025+0447}{PC\,0025+0447}{PC\,0025+0447}
\MakeObj{tvlm513-46546}{TVLM\%20513-46546}{TVLM\,513-46546}{TVLM\,513-46546}{TVLM\,513-46546}
\MakeObj{vb10}{VB\%2010}{VB\,10}{VB\,10}{VB\,10}
\MakeObj{vb8}{VB\%208}{VB\,8}{VB\,8}{VB\,8}
\MakeObj{yzcmi}{YZ\%20CMi}{YZ\,CMi}{YZ\,CMi}{YZ\,CMi}

\begin{document}

\title{Trends in Ultracool Dwarf Magnetism. II. The Inverse Correlation \\
  Between X-Ray Activity and Rotation as Evidence for a Bimodal Dynamo}
\author{
  B. A. Cook\altaffilmark{1},
  P. K. G. Williams\altaffilmark{2},
  and
  E. Berger\altaffilmark{2}
}
\email{bacook@princeton.edu}
\altaffiltext{1}{Department of Astrophysical Sciences, Princeton
  University, Princeton, NJ 08544, USA.}
\altaffiltext{2}{Harvard-Smithsonian Center for Astrophysics, 60
  Garden St., Cambridge, MA 02138, USA.}

\slugcomment{Draft: \today}
\shorttitle{UCD Magnetism Trends. II. X-Ray Activity, Rotation, and a Bimodal Dynamo}
\shortauthors{Cook \textit{et al.}}

\begin{abstract}
  Observations of magnetic activity indicators in solar-type stars
  exhibit a relationship with rotation with an increase until a
  ``saturation'' level and a moderate decrease in activity in the very
  fastest rotators (``supersaturation''). While X-ray data have
  suggested that this relationship is strongly violated in ultracool
  dwarfs (UCDs; spectral type $\gtrsim$M7), the limited number of
  X-ray detections has prevented firm conclusions. In this paper, we
  analyze the X-ray activity-rotation relation in 38 ultracool
  dwarfs. Our sample represents the largest catalog of X-ray active
  ultracool dwarfs to date, including seven new and four
  previously-unpublished \chandra\ observations presented in a
  companion paper. We identify a substantial number of
  rapidly-rotating UCDs with X-ray activity extending two orders of
  magnitude below the expected saturation level and measure a
  ``supersaturation''-type anticorrelation between rotation and X-ray
  activity. The scatter in UCD X-ray activity at a fixed rotation is
  \apx3 times larger than that in earlier-type stars. We discuss
  several mechanisms that have been proposed to explain the data,
  including centrifugal stripping of the corona, and find them to be
  inconsistent with the observed trends. Instead, we suggest that an
  additional parameter correlated with both X-ray activity and
  rotation is responsible for the observed effects. Building on the
  results of Zeeman-Doppler imaging of UCD magnetic fields and our
  companion study of radio/X-ray flux ratios, we argue that this
  parameter is the magnetic field topology, and that the large scatter
  in UCD X-ray fluxes reflects the presence of two dynamo modes that
  produce distinct topologies.
\end{abstract}

\keywords{stars: low mass --- stars: magnetic fields --- stars:
  rotation --- stars: activity}

\section{Introduction}
\label{s.intro}

Magnetic field generation in the Sun relies on differential rotation
in the radiative-convective boundary, and it is therefore expected
that this $\alpha\Omega$ dynamo \citep{p55} will not be responsible
for the magnetic fields of stars that are fully convective
\citep[dwarfs with spectral types $\gsim$M3, or masses
  $\lsim$$0.35$\Msun;][]{Chabrier1997}. It was thus originally
anticipated that magnetic activity would nont be found in the
``ultracool dwarfs'' (UCDs), very-low-mass stars and brown dwarfs with
spectral types $\gtrsim$M7 \citep{krl+99,mdb+99}. X-ray and
\ha\ emission, which generally trace magnetic activity, are indeed
significantly suppressed in these objects
\citep{bm95,mb03,gmr+00,mb03,whw+04,smf+06,rb08,bbf+10}.  However, the
first detection of radio emission, another tracer of magnetism, from a
brown dwarf by \citet{Berger2001} affirmed that UCDs can, in fact,
generate significant magnetic fields, and this result has been
confirmed by subsequent observations \citep{b02, berger2006, had+06,
  rb07, opbh+09, bbf+10, rb10, mdp+10, ahd+13}.

Rotation plays a key role in many stellar dynamo models, including the
$\alpha\Omega$ dynamo, and studies of the relationship between
rotation and magnetic activity therefore shed light on the dynamo
process. Solar-type stars obey a well-known relationship in which
rapid rotation leads to increasing X-ray and \ha\ emission relative to
\Lb, the stellar bolometric luminosity, up to a ``saturation''
level. In the X-ray band, the saturation level is $\LxLb \approx
10^{-3}$, and it is reached at rotation periods $\prot \lsim 2$--3
days \citeeg{Pizzolato2003,Wright2011}. These data are generally taken
to support a rotationally-powered dynamo with saturation possibly
originating in saturation of the dynamo itself \citep{vw87},
centrifugal stripping of the corona \citep{Jardine1999}, or filling of the
entire stellar surface with magnetic active regions \citep{v84}. The
same overall relationship is observed in early- and mid-M~dwarfs
\citep{dfpm98,mb03}. There is also evidence for a ``supersaturation''
effect in which the fastest rotators (projected rotational velocity
$\vsi \gsim 100$~\kms) have activity levels that are depressed by
factors of \apx2--3 \citep[$\LxLb \approx
  10^{-3.5}$;][]{Randich1996,Stepien2001,Wright2011}. Supersaturation
is not observed in \ha\ \citep{mcd09}.

Despite the presumed cessation of the $\alpha\Omega$ dynamo at the
transition to full convection, solar-type activity-rotation relations
are seen in mid-M~dwarfs \citep{dfpm98,mb03}. New phenomena begin to
appear, however, in the UCD regime. UCDs exhibit an enhanced
supersaturation-like effect in X-rays and \ha, with \LxLb\ and
\LhLb\ decreasing by 1--2 orders of magnitude at the highest
rotational velocities \citep{James2000,bgg+08,rb10}. UCD radio emission,
on the other hand, \emph{increases} with rotation, with no indications
of saturation in even the fastest rotators \citep{mbr12}. The
interpretation of these results is complicated by the number of other
changes occurring in this regime, such as the aforementioned general
dropoff of UCD coronal and chromospheric emissions, which is possibly
due to decoupling of the increasingly neutral stellar atmospheres from
the magnetic fields \citep{Mohanty2002} or more efficient trapping of
energetic electrons \citep{bgg+08,bbf+10}. Furthermore, the fully
convective nature of UCDs suggests that a separate dynamo mechanism is
in operation, which could quite plausibly have a different dependence
on rotation than the $\alpha\Omega$ dynamo \citeeg{Durney1993}.

In this paper, we build on the observations and database described in
\citet[hereafter \paperone]{paper1} to investigate the X-ray
activity-rotation relation in ultracool dwarfs. We assemble a
comprehensive sample of UCDs with both X-ray and rotation
measurements, including the new measurements presented in
\paperone. We proceed by reviewing the sample (\S\ref{s.obs}) and our
computation of various stellar parameters (\S\ref{s.stellarprops}). We
then analyze the rotation-activity relation in the UCD sample and
examine correlations between activity and various rotation parameters
(\S\ref{s.trends}). We discuss several mechanisms that have been
suggested to explain a decline in activity, comparing their
predictions to the observed relations in UCDs, and propose that
changes in UCD magnetic topology drive the observed anticorrelation
between rotation and X-ray activity (\S\ref{s.disc}). Finally, we
review our results and suggest future studies that can test our model
(\S\ref{s.conc}).

Throughout this work, we use the notation $[x] \equiv \log_{10} x$,
with $x$ being measured in cgs units if it is a diminsional quantity,
unless its units are specified otherwise.

\section{Observational Data}
\label{s.obs}

In \paperone\ we described observations of seven UCDs using both \chandra\ and
the Karl G. Jansky Very Large Array (VLA). These UCDs were chosen to span a
narrow spectral type range but a broad range of projected rotational
velocities. All seven targets were detected in the X-ray, while only
\obj{lhs2397} was detected in the radio. These new observations nearly double
the number of UCDs with X-ray detections. \paperone\ also describes our
analysis of archival \chandra\ observations of four sources, \objl{g208-44}
(M5.5+M8.5), \objl{g208-45} (M6), \objl{lp349} (M8+M9), and \objl{dp0255}
(L8), all of which have measured rotation velocities. The first three of these
are detected while the last is not.

In \paperone\ we also combined our new measurements with a thorough
investigation of the literature to build a database of UCDs having both X-ray
and radio measurements. We have augmented this database with measurements of
\vsi\ from the literature. In Tables~\ref{t.dbsrcs} and \ref{t.dbdata} we
present the compiled data. Some objects having radio measurements but not
\vsi\ appear in \paperone\ but not in this work; others have \vsi\ measurements
but no radio data, and appear only in this work.

% TableBuilder table
\begin{deluxetable*}{r@{\,}lllr@{}lr@{}lr@{}lr@{}lr@{}l@{\,}lccc}
%custom preamble

%hardcoded preamble
\tablecolumns{18}
\tablewidth{0em}
\tablecaption{UCDs with X-ray and \vsi\ measurements\label{t.dbsrcs}}
\tablehead{
\colhead{} & \colhead{2MASS Identifier} & \colhead{Other Name} & \colhead{SpT} & \multicolumn{2}{c}{$J$} & \multicolumn{2}{c}{$K_s$} & \multicolumn{2}{c}{$d$} & \multicolumn{2}{c}{[$\Lb$]} & \multicolumn{3}{c}{\vsi} & \multicolumn{3}{c}{References}   \\
 \cline{16-18}  &  &  &  & \multicolumn{2}{c}{(mag)} & \multicolumn{2}{c}{(mag)} & \multicolumn{2}{c}{(pc)} & \multicolumn{2}{c}{[$L_\sun$]} & \multicolumn{3}{c}{(\kms)} &  &  &  \\ \\
\multicolumn{1}{c}{} & \multicolumn{1}{c}{(1)} & \multicolumn{1}{c}{(2)} & \multicolumn{1}{c}{(3)} & \multicolumn{2}{c}{(4)} & \multicolumn{2}{c}{(5)} & \multicolumn{2}{c}{(6)} & \multicolumn{2}{c}{(7)} & \multicolumn{3}{c}{(8)} & \multicolumn{1}{c}{(S)} & \multicolumn{1}{c}{(D)} & \multicolumn{1}{c}{(V)}
}
\startdata
 & \object{08294949$+$2646348} & LHS 248 & M6.5 & $8$ & $.23$ & $7$ & $.26$ & $3$ & $.6$ & $-3$ & $.09$ & $11$ & $.0$  & $\pm\,2$ & 1 & 2 & 3 \\
\P & \object{10481258$-$1120082} & LHS 292 & M6.5 & $8$ & $.86$ & $7$ & $.93$ & $4$ & $.5$ & $-3$ & $.15$ & \multicolumn{3}{c}{$<$$3$} & 1 & 4 & 5 \\
\P & \object{22285440$-$1325178} & GJ 4281 & M6.5 & $10$ & $.77$ & $9$ & $.84$ & $11$ & $.3$ & $-3$ & $.13$ & $7$ & $.0$  & $\pm\,2$ & 6 & 4 & 3 \\
 & \object{11571691$+$2755489} & CTI 115638.4$+$280000 & M7 & $14$ & $.32$ & $13$ & $.34$ & $49$ & $.1$ & $-3$ & $.26$ & $10$ & $.5$  & $\pm\,2$ & 7 & 8 & 3 \\
 & \object{13142039$+$1320011 AB} & NLTT 33370 AB & M7 & $9$ & $.75$ & $8$ & $.79$ & $16$ & $.4$ & $-2$ & $.40$ & $45$ &  & $\pm\,5$ & 9 & 9 & 10 \\
 & \object{14563831$-$2809473} & LHS 3003 & M7 & $9$ & $.96$ & $8$ & $.93$ & $6$ & $.4$ & $-3$ & $.29$ & $5$ &  & $\pm\,2$ & 11 & 4 & 5 \\
 & \object{16553529$-$0823401} & vB 8 & M7 & $9$ & $.78$ & $8$ & $.82$ & $6$ & $.5$ & $-3$ & $.21$ & $9$ & $.0$  & $\pm\,2$ & 1 & 12 & 3 \\
\P & \object{11554286$-$2224586} & LP 851$-$346 & M7.5 & $10$ & $.93$ & $9$ & $.88$ & $9$ & $.7$ & $-3$ & $.32$ & $33$ &  & $\pm\,3$ & 13 & 13 & 5 \\
\P & \object{15210103$+$5053230} & NLTT 40026 & M7.5 & $12$ & $.01$ & $10$ & $.92$ & $16$ & $.2$ & $-3$ & $.30$ & $40$ &  & $\pm\,4$ & 11 & 11 & 5 \\
\P & \object{00275592$+$2219328 AB} & LP 349$-$25 AB & M8 & $10$ & $.61$ & $9$ & $.57$ & $13$ & $.2$ & $-2$ & $.93$ & $56$ & $.0$  & $\pm\,6.0$ & 11 & 14 & 15 \\
 & \object{03205965$+$1854233} & LP 412$-$31 & M8 & $11$ & $.76$ & $10$ & $.64$ & $14$ & $.5$ & $-3$ & $.29$ & $15$ &  & $\pm\,4.5$ & 11 & 16 & 5 \\
\P & \object{11214924$-$1313084 AB} & LHS 2397a AB & M8 & $11$ & $.93$ & $10$ & $.73$ & $14$ & $.3$ & $-3$ & $.36$ & $20$ & $.0$  & $\pm\,2$ & 17 & 4 & 3 \\
\P & \object{18432213$+$4040209} & LHS 3406 & M8 & $11$ & $.31$ & $10$ & $.31$ & $14$ & $.1$ & $-3$ & $.16$ & $5$ &  & $\pm\,3.2$ & 11 & 4 & 5 \\
 & \object{19165762$+$0509021} & vB 10 & M8 & $9$ & $.91$ & $8$ & $.77$ & $6$ & $.1$ & $-3$ & $.30$ & $6$ & $.5$  & $\pm\,2$ & 18 & 18 & 3 \\
 & \object{01273917$+$2805536} & CTI 012657.5$+$280202 & M8.5 & $14$ & $.04$ & $12$ & $.86$ & $34$ & $.8$ & $-3$ & $.44$ & $11$ & $.0$  & $\pm\,2$ & 11 & 11 & 3 \\
 & \object{07075327$-$4900503} &  & M8.5 & $13$ & $.23$ & $12$ & $.11$ & $15$ & $.1$ & $-3$ & $.85$ & $10$ & $.0$  & $\pm\,2$ & 19 & 4 & 3 \\
 & \object{14542923$+$1606039 Ba} & Gl 569 Ba & M8.5 & $11$ & $.14$ & $10$ & $.02$ & $9$ & $.8$ & $-3$ & $.39$ & $19$ &  & $\pm\,2$ & 20 & 21 & 22 \\
 & \object{18353790$+$3259545} & LSPM J1835$+$3259 & M8.5 & $10$ & $.27$ & $9$ & $.17$ & $5$ & $.7$ & $-3$ & $.52$ & $44$ &  & $\pm\,4$ & 11 & 16 & 5 \\
\P & \object{01095117$-$0343264} & LP 647$-$13 & M9 & $11$ & $.69$ & $10$ & $.43$ & $11$ & $.1$ & $-3$ & $.48$ & $13$ &  & $\pm\,2$ & 11 & 11 & 5 \\
 & \object{03393521$-$3525440} & LP 944$-$20 & M9 & $10$ & $.72$ & $9$ & $.55$ & $5$ & $.0$ & $-3$ & $.81$ & $26$ &  & $\pm\,3$ & 11 & 16 & 5 \\
 & \object{08533619$-$0329321} & LHS 2065 & M9 & $11$ & $.21$ & $9$ & $.94$ & $8$ & $.5$ & $-3$ & $.52$ & $13$ & $.5$  & $\pm\,2$ & 11 & 4 & 5 \\
 & \object{10481463$-$3956062} &  & M9 & $9$ & $.54$ & $8$ & $.45$ & $4$ & $.0$ & $-3$ & $.54$ & $18$ &  & $\pm\,2$ & 5 & 19 & 5 \\
 & \object{11592743$-$5247188} & 1RXS J115928.5$-$524717 & M9 & $11$ & $.43$ & $10$ & $.32$ & $10$ & $.1$ & $-3$ & $.48$ & \multicolumn{3}{c}{$\sim$$25$} & 23 & 24 & 24 \\
 & \object{14284323$+$3310391} & LHS 2924 & M9 & $11$ & $.99$ & $10$ & $.74$ & $10$ & $.8$ & $-3$ & $.63$ & $11$ & $.0$  & $\pm\,2$ & 19 & 4 & 3 \\
 & \object{14542923$+$1606039 Bb} & Gl 569 Bb & M9 & $11$ & $.65$ & $10$ & $.43$ & $9$ & $.8$ & $-3$ & $.58$ & $6$ &  & $\pm\,3$ & 20 & 21 & 22 \\
 & \object{15010818$+$2250020} & TVLM 513$-$46546 & M9 & $11$ & $.87$ & $10$ & $.71$ & $9$ & $.9$ & $-3$ & $.67$ & $60$ & $.0$  & $\pm\,2$ & 19 & 4 & 3 \\
 & \object{00242463$-$0158201} & BRI B0021$-$0214 & M9.5 & $11$ & $.99$ & $10$ & $.54$ & $12$ & $.1$ & $-3$ & $.49$ & $33$ &  & $\pm\,3$ & 25 & 4 & 5 \\
 & \object{00274197$+$0503417} & PC 0025$+$0447 & M9.5 & $16$ & $.19$ & $14$ & $.96$ & $72$ & $.0$ & $-3$ & $.67$ & $13$ &  & $\pm\,3$ & 26 & 26 & 27 \\
 & \object{07464256$+$2000321 AB} &  & L0 & $11$ & $.76$ & $10$ & $.47$ & $12$ & $.2$ & $-3$ & $.36$ & $31$ &  & $\pm\,3$ & 28 & 11 & 29 \\
 & \object{06023045$+$3910592} & LSR J0602$+$3910 & L1 & $12$ & $.30$ & $10$ & $.87$ & $10$ & $.6$ & $-3$ & $.67$ & $9$ &  & $\pm\,3$ & 30 & 30 & 29 \\
 & \object{13054019$-$2541059 AB} & Kelu$-$1 AB & L2 & $13$ & $.41$ & $11$ & $.75$ & $18$ & $.7$ & $-3$ & $.56$ & $76$ &  & $\pm\,8$ & 11 & 11 & 29 \\
 & \object{05233822$-$1403022} &  & L2.5 & $13$ & $.08$ & $11$ & $.64$ & $13$ & $.4$ & $-3$ & $.82$ & $21$ &  & $\pm\,3$ & 11 & 16 & 29 \\
 & \object{00361617$+$1821104} & LSPM J0036$+$1821 & L3.5 & $12$ & $.47$ & $11$ & $.06$ & $8$ & $.8$ & $-3$ & $.99$ & $45$ &  & $\pm\,5$ & 31 & 19 & 29 \\
 & \object{14501581$+$2354424 B} & HD 130948 B & L4 & $13$ & $.90$ & $13$ & $.30$ & $17$ & $.9$ & $-4$ & $.28$ & $62$ &  & $\pm\,4$ & 32 & 33 & 22 \\
 & \object{14501581$+$2354424 C} & HD 130948 C & L4 & $14$ & $.20$ & $12$ & $.60$ & $17$ & $.9$ & $-4$ & $.00$ & $86$ &  & $\pm\,6$ & 32 & 33 & 22 \\
 & \object{12281523$-$1547342 AB} &  & L5 & $14$ & $.38$ & $12$ & $.77$ & $20$ & $.2$ & $-3$ & $.93$ & $22$ & $.0$  & $\pm\,2$ & 11 & 11 & 3 \\
 & \object{15074769$-$1627386} &  & L5 & $12$ & $.83$ & $11$ & $.31$ & $7$ & $.3$ & $-4$ & $.23$ & $32$ &  & $\pm\,3$ & 31 & 16 & 29 \\
\P & \object{02550357$-$4700509} &  & L8 & $13$ & $.25$ & $11$ & $.56$ & $5$ & $.0$ & $-4$ & $.58$ & $67$ &  & $\pm\,13$ & 11 & 34 & 29
\enddata
\tablecomments{Rows marked with a pilcrow (\P) indicate sources with new measurements
presented in \paperone. Col. (3) is spectral type. Cols. (4) and (5) 2MASS
\citep{the2mass}.}
\tablerefs{Columns are (S), spectral type; (D), distance; and (V), \vsi. [1] \citet{hks94}, [2] \citet{sl04}, [3] \citet{mb03}, [4] \citet{thegctp}, [5] \citet{rb10}, [6] \citet{khm91}, [7] \citet{fgsb93}, [8] \citet{khs95}, [9] \citet{ltsr09}, [10] \citet{mbi+11}, [11] \citet{crl+03}, [12] \citet{thethirdgj}, [13] \citet{cpbd+05}, [14] \citet{gc09}, [15] \citet{dmm+12}, [16] \citet{crk+07}, [17] \citet{fcs03}, [18] \citet{bbg+08}, [19] \citet{rck+08}, [20] \citet{lzob+01}, [21] \citet{s04}, [22] \citet{kgf+12}, [23] \citet{hss+04}, [24] \citet{rs09}, [25] \citet{rhg95}, [26] \citet{mbr12}, [27] \citet{mbzo99}, [28] \citet{brpb+09}, [29] \citet{rb08}, [30] \citet{bbf+10}, [31] \citet{brr+05}, [32] \citet{smf+06}, [33] \citet{pmc+02}, [34] \citet{cmj+06}}
\end{deluxetable*}
% end TableBuilder table

% TableBuilder table
\begin{deluxetable*}{llr@{}lc@{\,}cr@{\,}r@{}lr@{\,}r@{}lc}
%custom preamble

%hardcoded preamble
\tablecolumns{13}
\tablewidth{0em}
\tablecaption{X-ray data for UCDs with \vsi\ measurements\label{t.dbdata}}
\tablehead{
\colhead{2MASS Identifier} & \colhead{SpT} & \multicolumn{2}{c}{[$\Lb$]} & \colhead{St.} & \colhead{X-Ray Band} & \multicolumn{3}{c}{[$L_X$]} & \multicolumn{3}{c}{[$L_X/\Lb$]} & \colhead{Ref.} \\
 &  & \multicolumn{2}{c}{[$L_\sun$]} &  & \colhead{(keV)} & \multicolumn{3}{c}{[erg s$^{-1}$]} &  &  &  &  \\ \\
\multicolumn{1}{c}{(1)} & \multicolumn{1}{c}{(2)} & \multicolumn{2}{c}{(3)} & \multicolumn{1}{c}{(4)} & \multicolumn{1}{c}{(5)} & \multicolumn{3}{c}{(6)} & \multicolumn{3}{c}{(7)} & \multicolumn{1}{c}{(8)}
}
\startdata
\object{08294949$+$2646348} & M6.5 & $-3$ & $.09$ & -- & $0.1$--$2.4$ &  & $26$ & $.5$ &  & $-4$ & $.0$ & 1 \\
\object{10481258$-$1120082} & M6.5 & $-3$ & $.15$ & -- & $0.2$--$2.0$ &  & $25$ & $.7$ &  & $-4$ & $.7$ & $\star$ \\
\object{22285440$-$1325178} & M6.5 & $-3$ & $.13$ & -- & $0.2$--$2.0$ &  & $25$ & $.7$ &  & $-4$ & $.8$ & $\star$ \\
\object{11571691$+$2755489} & M7 & $-3$ & $.26$ & -- & $0.1$--$2.4$ & $<$ & $28$ & $.3$ & $<$ & $-2$ & $.0$ & 4 \\
\object{13142039$+$1320011 AB} & M7 & $-2$ & $.40$ & Q & $0.2$--$2.0$ &  & $27$ & $.6$ &  & $-3$ & $.6$ & 5 \\
 &  &  &  & F & $0.2$--$2.0$ &  & $28$ & $.3$ &  & $-2$ & $.9$ & 5 \\
\object{14563831$-$2809473} & M7 & $-3$ & $.29$ & -- & $0.1$--$2.4$ &  & $26$ & $.2$ &  & $-4$ & $.1$ & 7 \\
\object{16553529$-$0823401} & M7 & $-3$ & $.21$ & Q & $0.1$--$2.4$ &  & $26$ & $.9$ &  & $-3$ & $.5$ & 8 \\
 &  &  &  & F & $0.1$--$2.4$ &  & $27$ & $.5$ &  & $-2$ & $.9$ & 7 \\
\object{11554286$-$2224586} & M7.5 & $-3$ & $.32$ & Q & $0.2$--$2.0$ &  & $25$ & $.9$ &  & $-4$ & $.4$ & $\star$ \\
 &  &  &  & F & $0.2$--$2.0$ &  & $26$ & $.8$ &  & $-3$ & $.5$ & $\star$ \\
\object{15210103$+$5053230} & M7.5 & $-3$ & $.30$ & -- & $0.2$--$2.0$ &  & $25$ & $.7$ &  & $-4$ & $.6$ & $\star$ \\
\object{00275592$+$2219328 AB} & M8 & $-2$ & $.93$ & Q & $0.2$--$2.0$ &  & $25$ & $.8$ &  & $-4$ & $.8$ & $\star$ \\
 &  &  &  & F & $0.2$--$2.0$ &  & $26$ & $.3$ &  & $-4$ & $.3$ & $\star$ \\
\object{03205965$+$1854233} & M8 & $-3$ & $.29$ & Q & $0.3$--$8.0$ &  & $27$ & $.2$ &  & $-3$ & $.1$ & 10 \\
 &  &  &  & F & $0.3$--$8.0$ &  & $29$ & $.7$ &  & $-0$ & $.6$ & 10 \\
\object{11214924$-$1313084 AB} & M8 & $-3$ & $.36$ & -- & $0.2$--$2.0$ &  & $27$ & $.1$ &  & $-3$ & $.1$ & $\star$ \\
\object{18432213$+$4040209} & M8 & $-3$ & $.16$ & Q & $0.2$--$2.0$ &  & $26$ & $.5$ &  & $-4$ & $.0$ & $\star$ \\
 &  &  &  & F & $0.2$--$2.0$ &  & $27$ & $.1$ &  & $-3$ & $.3$ & $\star$ \\
\object{19165762$+$0509021} & M8 & $-3$ & $.30$ & Q & $0.2$--$2.0$ &  & $25$ & $.2$ &  & $-5$ & $.1$ & 11 \\
 &  &  &  & F & $0.2$--$2.0$ &  & $25$ & $.7$ &  & $-4$ & $.6$ & 11 \\
\object{01273917$+$2805536} & M8.5 & $-3$ & $.44$ & -- & $0.1$--$2.4$ & $<$ & $27$ & $.8$ & $<$ & $-2$ & $.3$ & 4 \\
\object{07075327$-$4900503} & M8.5 & $-3$ & $.85$ & -- & $0.1$--$2.4$ & $<$ & $27$ & $.4$ & $<$ & $-2$ & $.3$ & 4 \\
\object{14542923$+$1606039 Ba} & M8.5 & $-3$ & $.39$ & -- & $0.5$--$8.0$ &  & $26$ & $.9$ &  & $-3$ & $.3$ & 12 \\
\object{18353790$+$3259545} & M8.5 & $-3$ & $.52$ & -- & $0.2$--$2.0$ & $<$ & $24$ & $.5$ & $<$ & $-5$ & $.6$ & 11 \\
\object{01095117$-$0343264} & M9 & $-3$ & $.48$ & Q & $0.2$--$2.0$ &  & $25$ & $.2$ &  & $-4$ & $.9$ & $\star$ \\
 &  &  &  & F & $0.2$--$2.0$ &  & $26$ & $.1$ &  & $-4$ & $.0$ & $\star$ \\
\object{03393521$-$3525440} & M9 & $-3$ & $.81$ & Q & $0.1$--$10.0$ & $<$ & $24$ & $.0$ & $<$ & $-5$ & $.7$ & 14 \\
 &  &  &  & F & $0.1$--$10.0$ &  & $25$ & $.6$ &  & $-4$ & $.2$ & 14 \\
\object{08533619$-$0329321} & M9 & $-3$ & $.52$ & Q & $0.3$--$0.8$ &  & $26$ & $.5$ &  & $-3$ & $.6$ & 15 \\
 &  &  &  & F & $0.1$--$2.4$ &  & $27$ & $.6$ &  & $-2$ & $.5$ & 16 \\
\object{10481463$-$3956062} & M9 & $-3$ & $.54$ & -- & $0.2$--$2.0$ &  & $25$ & $.1$ &  & $-4$ & $.9$ & 17 \\
\object{11592743$-$5247188} & M9 & $-3$ & $.48$ & Q & $0.1$--$2.4$ & $<$ & $27$ & $.9$ & $<$ & $-2$ & $.2$ & 18 \\
 &  &  &  & F & $0.1$--$2.4$ &  & $28$ & $.8$ &  & $-1$ & $.3$ & 18 \\
\object{14284323$+$3310391} & M9 & $-3$ & $.63$ & -- & $0.1$--$2.4$ & $<$ & $25$ & $.5$ & $<$ & $-4$ & $.4$ & 4 \\
\object{14542923$+$1606039 Bb} & M9 & $-3$ & $.58$ & -- & $0.5$--$8.0$ &  & $26$ & $.9$ &  & $-3$ & $.1$ & 12 \\
\object{15010818$+$2250020} & M9 & $-3$ & $.67$ & Q & $0.3$--$2.0$ &  & $24$ & $.9$ &  & $-5$ & $.0$ & 20 \\
 &  &  &  & F & $0.3$--$2.0$ &  & $27$ & $.3$ &  & $-2$ & $.6$ & 20 \\
\object{00242463$-$0158201} & M9.5 & $-3$ & $.49$ & -- & $0.2$--$2.0$ & $<$ & $25$ & $.1$ & $<$ & $-4$ & $.9$ & 21 \\
\object{00274197$+$0503417} & M9.5 & $-3$ & $.67$ & -- & $0.3$--$8.0$ & $<$ & $26$ & $.2$ & $<$ & $-3$ & $.7$ & 17 \\
\object{07464256$+$2000321 AB} & L0 & $-3$ & $.36$ & -- & $0.2$--$2.0$ & $<$ & $25$ & $.2$ & $<$ & $-5$ & $.0$ & 23 \\
\object{06023045$+$3910592} & L1 & $-3$ & $.67$ & -- & $0.2$--$2.0$ & $<$ & $25$ & $.0$ & $<$ & $-4$ & $.9$ & 21 \\
\object{13054019$-$2541059 AB} & L2 & $-3$ & $.56$ & -- & $0.1$--$10.0$ &  & $25$ & $.3$ &  & $-4$ & $.7$ & 25 \\
\object{05233822$-$1403022} & L2.5 & $-3$ & $.82$ & -- & $0.2$--$2.0$ & $<$ & $25$ & $.2$ & $<$ & $-4$ & $.5$ & 21 \\
\object{00361617$+$1821104} & L3.5 & $-3$ & $.99$ & -- & $0.2$--$8.0$ & $<$ & $24$ & $.9$ & $<$ & $-4$ & $.7$ & 26 \\
\object{14501581$+$2354424 B} & L4 & $-4$ & $.28$ & -- & $0.5$--$8.0$ & $<$ & $25$ & $.7$ & $<$ & $-3$ & $.6$ & 12 \\
\object{14501581$+$2354424 C} & L4 & $-4$ & $.00$ & -- & $0.5$--$8.0$ & $<$ & $25$ & $.7$ & $<$ & $-3$ & $.9$ & 12 \\
\object{12281523$-$1547342 AB} & L5 & $-3$ & $.93$ & -- & $0.3$--$8.0$ & $<$ & $26$ & $.6$ & $<$ & $-3$ & $.0$ & 17 \\
\object{15074769$-$1627386} & L5 & $-4$ & $.23$ & -- & $0.2$--$8.0$ & $<$ & $24$ & $.8$ & $<$ & $-4$ & $.6$ & 26 \\
\object{02550357$-$4700509} & L8 & $-4$ & $.58$ & -- & $0.2$--$2.0$ & $<$ & $24$ & $.3$ & $<$ & $-4$ & $.7$ & $\star$
\enddata
\tablecomments{Col. (4) is the state of the source: quiescent (Q), flaring (F), or
indeterminate/unknown (--). \Lx\ has been normalized to the 0.2--2~keV
bandpass as described in the text.}
\tablerefs{Col. (8) is the X-ray luminosity reference. [$\star$]: \paperone, [1] \citet{dfpm98}, [2] \citet{mb03}, [3] \citet{rb10}, [4] \citet{fgsb93}, [5] \citet[in preparation]{dracut}, [6] \citet{mbi+11}, [7] \citet{sfg95}, [8] \citet{gsbf93}, [9] \citet{dmm+12}, [10] \citet{ssml06}, [11] \citet{bbg+08}, [12] \citet{smf+06}, [13] \citet{kgf+12}, [14] \citet{rbmb00}, [15] \citet{rs08}, [16] \citet{sl02}, [17] \citet{sab+12}, [18] \citet{hss+04}, [19] \citet{rs09}, [20] \citet{bgg+08}, [21] \citet{bbf+10}, [22] \citet{mbzo99}, [23] \citet{brpb+09}, [24] \citet{rb08}, [25] \citet{aob+07}, [26] \citet{brr+05}}
\end{deluxetable*}
% end TableBuilder table

\section{Stellar Rotation Properties}
\label{s.stellarprops}

Stellar X-ray activity is commonly analyzed in terms of the ratio of
X-ray to bolometric luminosity, thus indicating what fraction of the
total radiative energy is emitted in X-rays. We use the term
\textit{X-ray activity} to mean the ratio of X-ray luminosity (\Lx, in
0.2--2.0 keV band) to the bolometric luminosity (\Lb). We calculate
bolometric luminosities as described in the Appendix of \paperone; in
short, we use near-IR absolute magnitudes and bolometric corrections,
taking the $J$- and $K$-band corrections of \citet{Wilking1999} for
M~dwarfs and the $K$-band correction of \citet{Nakajima2004} for
L~dwarfs.

Projected rotation periods are derived from \vsi\ and stellar
radii. We approximate radii using the empirical mass-luminosity
relation of \citet{Delfosse2000} in conjunction with the theoretically
derived mass-radius relation of \citet{Baraffe1998}. This method is
consistent with \citet{rb10} and \citet{mbr12}, who studied the
relationship of \ha{} and radio with rotation, respectively. The stellar
radii are fairly constant in the UCD regime, and thus any
uncertainties in the rotational period are not dominated by the
mass-radius relation. However, we note that, due to the confounding
inclination parameter, the rotational periods derived from \vsi\ are
upper limits. X-ray emission is expected to be mostly isotropic, so
there should be no inherent correlation between \Lx\ and $\sin i$.

\citet{Noyes1984} first suggested that the relation between rotation
and magnetic activity may be better analyzed in terms of the Rossby
number $\ro \equiv \prot / \tauc$, where \tauc\ is the convective
turnover time. This proposition has been supported by subsequent work
\citeeg{Pizzolato2003, Wright2011}. We calculate \tauc\ using the
method of \citet{Kiraga2007} and \citet{rb10}, noting, however, that
the physical meaning of this quantity is ill-defined at low masses
\citeeg{kd96}:
\begin{equation}
\tauc (\text{days}) = \begin{cases}
  61.7 - 44.7 m, & 0.82 \leq m < 1.30 \\
  25,            & 0.65 \leq m < 0.82 \\
  86.9 - 94.3 m, & 0.10 \leq m < 0.65 \\
  70,            & m \leq 0.10
\end{cases}
\end{equation}
where $m = M_* / \Msun$. Because \tauc\ is taken to be constant for spectral
types later than \apx M7, the Rossby number is therefore essentially
equivalent to the rotation period for almost all of our sample of UCDs. The
use of \ro\ is nonetheless still important because it allows comparison across
a broad range of spectral types.

We also calculate the corotation radius, the radius at which the centripetal
force of corotation with the star is balanced by gravity:
\begin{equation}
\begin{split}
\rco / R_* &= \left(\frac{GM_*\prot^2}{4\pi^2R_*^3}\right)^{1/3} \\
 &\approx 41 m^{1/3}
  \left(\frac{\prot}{1\text{ day}}\right)^{2/3}
  \left(\frac{R_*}{R_J}\right)^{-1},
\end{split}
\end{equation}
with $R_J$ the radius of Jupiter. The ratio of the corotation radius to the stellar radius is often used as a
measure of the strength of centrifugal stripping effects in the corona
\citep{Jardine1999,James2000}.

\section{X-ray Activity Trends}
\label{s.trends}

\begin{figure*}[hbt]
\plotonebig{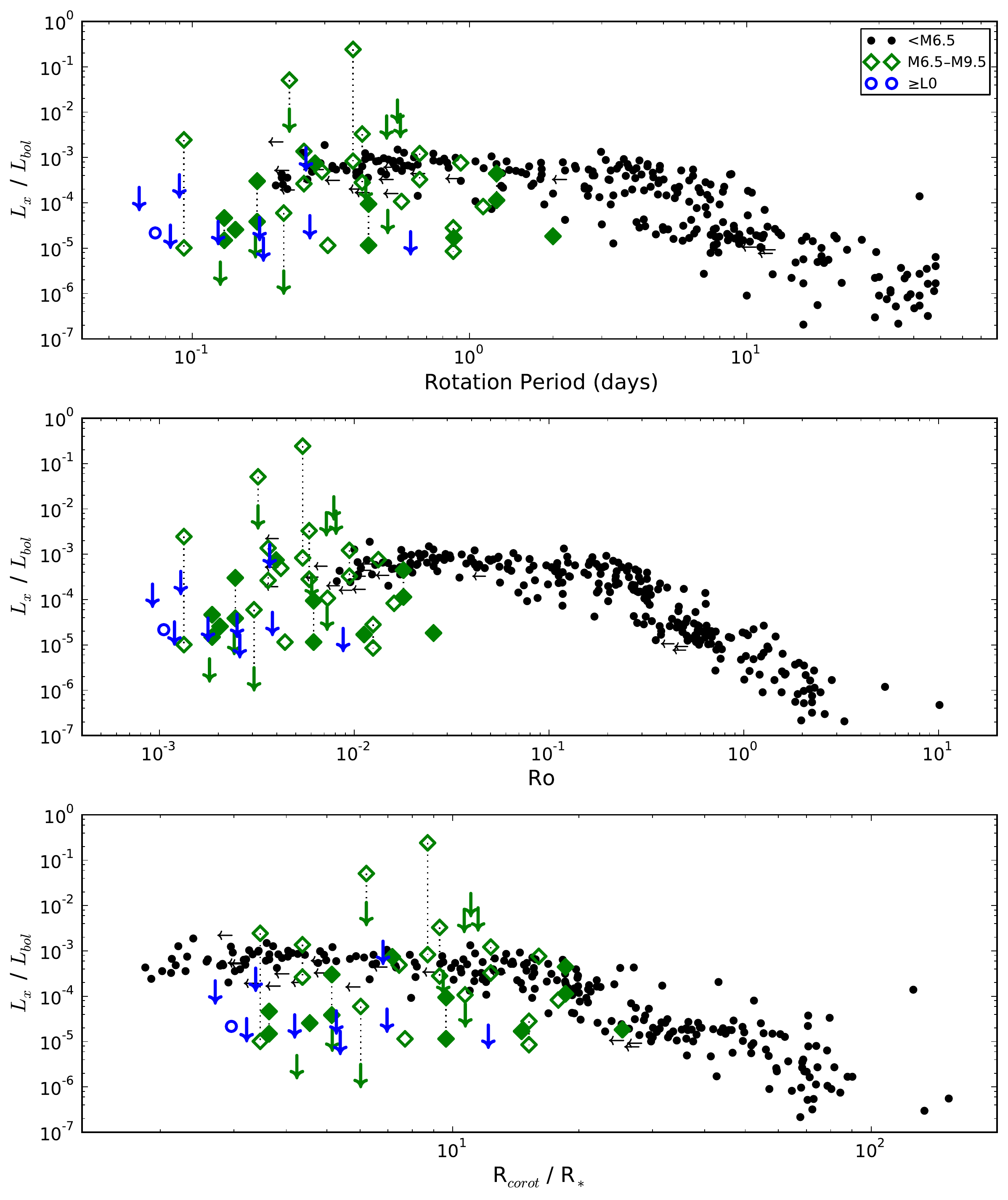}{.95}
\caption{X-ray activity versus various stellar rotation parameters, as
  calculated in \S\ref{s.stellarprops}. The parameters are rotation
  period (\textit{upper panel}), Rossby number (\emph{middle panel}),
  and the centrifugal stripping parameter \citep[\emph{lower
      panel}]{Jardine1999}. The UCD rotation parameters are formally
  upper limits due to the unknown inclination. Plotted are objects
  from this work, \citet{James2000}, and \citet{Pizzolato2003}. Points
  connected by a dotted line represent quiescent and flaring
  observations of the same object.  Spectral types are denoted by
  black circles ($<$M6.5), green diamonds (M6.5--M9.5), and blue
  circles ($\geq$L0). Downward-facing arrows indicate X-ray
  nondetections. Solid colored symbols indicate new measurements
  presented in \paperone. The $<$M6.5 outlier at $\prot \apx 40$~d and
  $\LxLb \apx 10^{-4}$ is Proxima Centauri, an unusually
  slowly-rotating M6 dwarf. In non-UCDs, a standard activity-rotation
  relation with saturation around $\LxLb \approx 10^{-3}$ is seen
  regardless of the rotation parameter used.  UCDs show rapid rotation
  but activity well below the saturation level. The fastest-rotating
  UCDs have larger centrifugal stripping parameters than many of the
  slower-rotating solar-type stars, suggesting that this effect is not
  responsible for ``supersaturation'' in UCDs (\S\ref{s.supersat}).}
\label{f.act-rot1}
\end{figure*}

\begin{figure*}[hbt]
\plotonebig{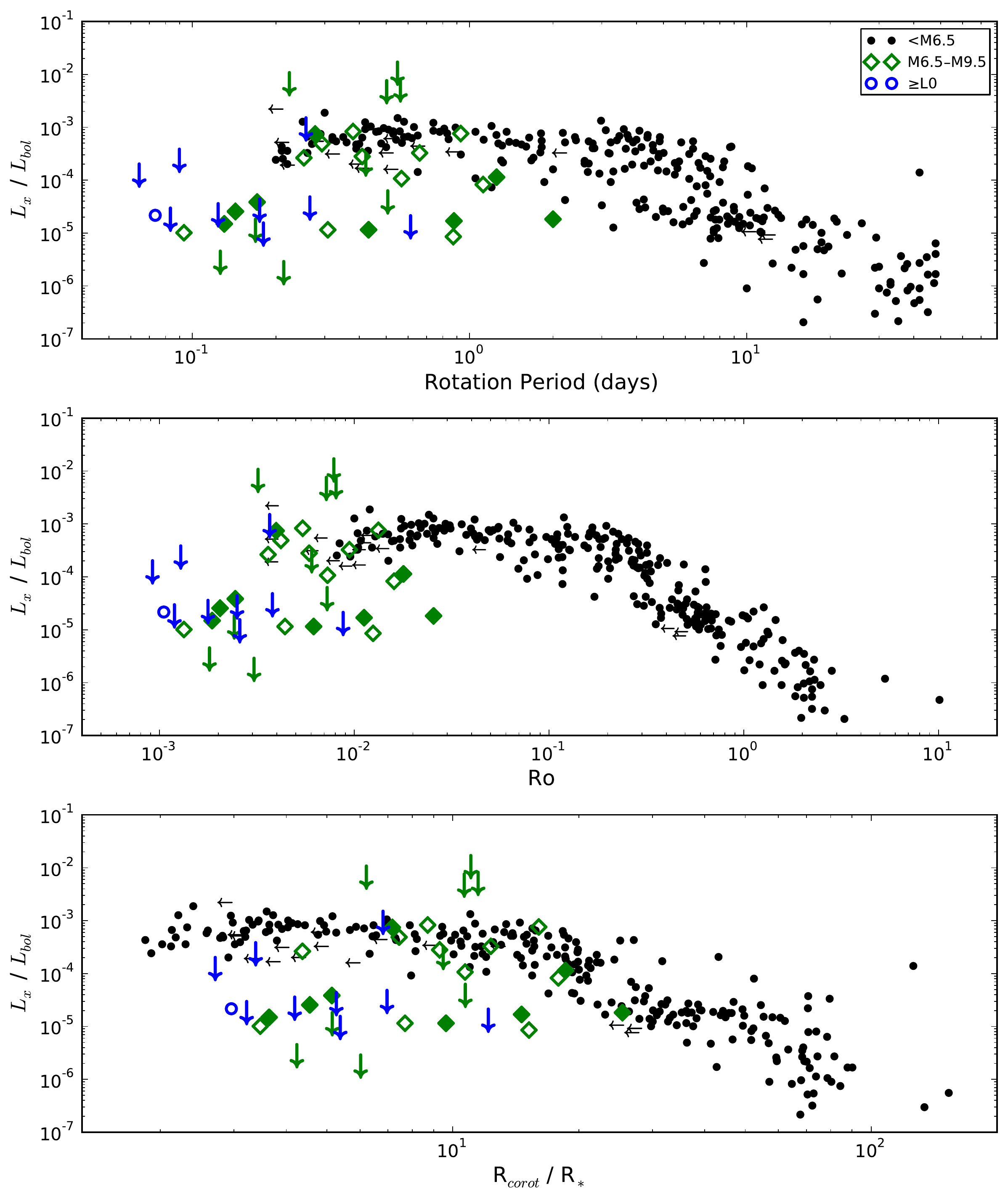}{.95}
\caption{The same as Figure~\ref{f.act-rot1}, but excluding known X-ray
  flares.}
\label{f.act-rot-nofl}
\end{figure*}

The X-ray activity of our ultracool dwarf sample is plotted against the
rotation period, Rossby number, and corotation radius in
Figure~\ref{f.act-rot1}. Also included for comparison are the samples of
\citet{Pizzolato2003} and \citet{James2000}, who studied the X-ray
activity-rotation relation in A--M6 and M0--M5 dwarfs, respectively. To
enhance clarity, we plot the same data excluding known X-ray flares in
Figure~\ref{f.act-rot-nofl}.

\begin{figure}[hbt]
\plotonebig{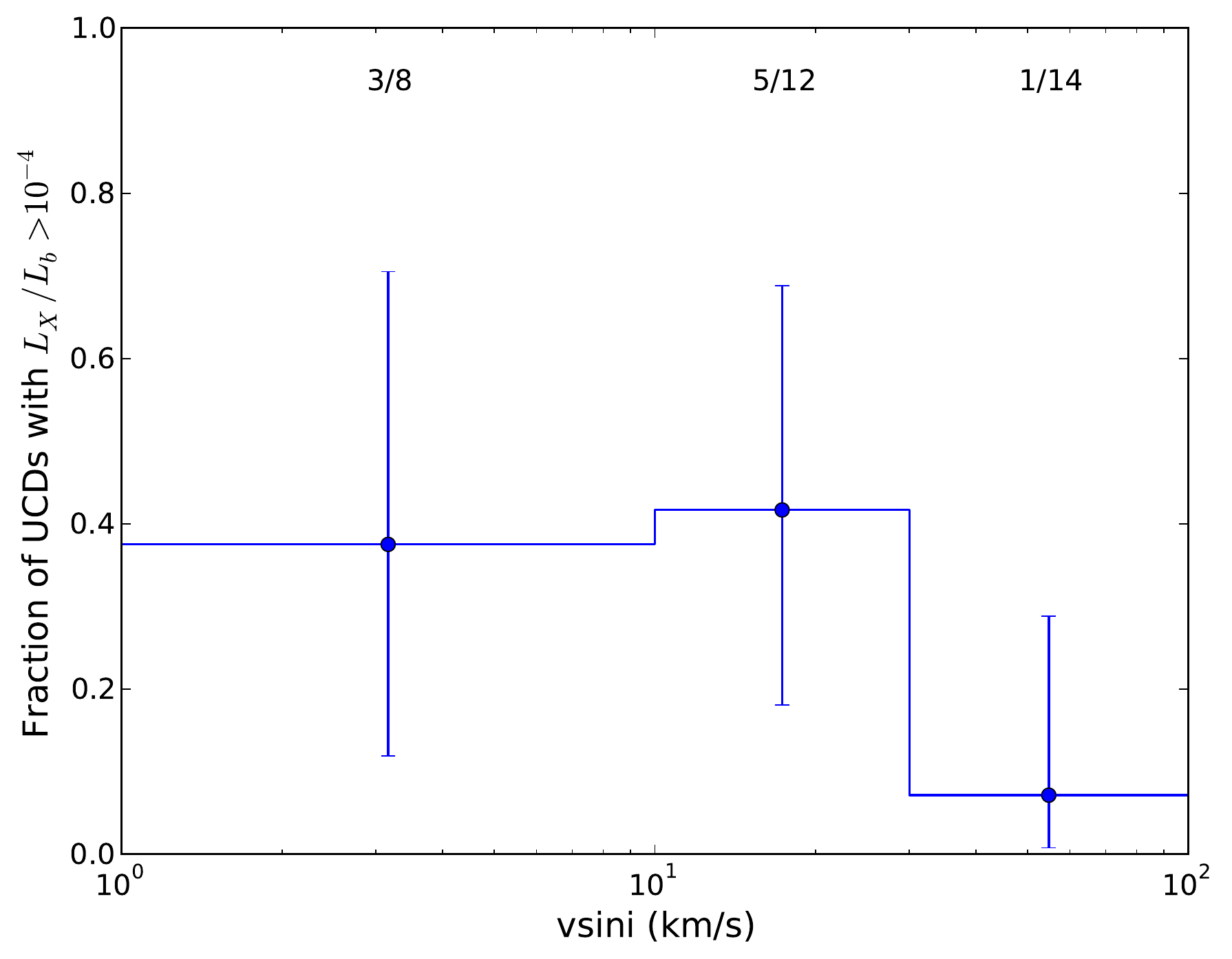}{0.95}
\caption{Fraction of UCDs with $\LxLb > 10^{-4}$ as a function of \vsi. Flares
  and upper limits less constraining than $\LxLb < 10^{-3}$ are not included
  in the analysis. The fractions above each bin give the actual counts. Error
  bars are 95\% confidence intervals for the binomial distribution computed
  using an equal-tailed Jeffreys interval.}
\label{f.detfrac}
\end{figure}

Figure~\ref{f.act-rot1} clearly shows the pre-saturation and
saturation regimes in stars earlier than type M6.5. UCDs are
systematically faster rotators than earlier-type stars in terms of
both period and Rossby number, having $P_\text{rot} < 2$~day and $\ro
< 0.03$. The UCDs do not follow the rotation-activity trend seen in
earlier-type stars, which would predict that their X-ray emission
should be saturated near $\LxLb \approx 10^{-3}$.  Instead, 22 of the
38 sources show X-ray emission at $\LxLb \le 10^{-4}$ and five have
$\LxLb \le 10^{-5}$. This dropoff is much stronger than that
associated with supersaturation in solar-type stars, in which
\LxLb\ drops only to \apx$10^{-3.5}$ \citep{Wright2011}. All L~dwarfs
(with the exception of one shallow non-detection) have activity levels
well below the saturation value.

The dropoff in UCD X-ray activity appears to be correlated with rotation. We
convey this effect in a simple way in Figure~\ref{f.detfrac}, which plots the
fraction of UCDs with $\LxLb > 10^{-4}$ when binned in terms of \vsi. Flaring
measurements and upper limits less constraining than $\LxLb < 10^{-3}$ are not
included in this analysis. There is a lack of relatively X-ray bright UCDs at
$\vsi \gtrsim 20$--30~\kms. Interestingly, \citet{mbr12} found that a large
fraction of UCDs above this cutoff are relatively radio-bright. In
\paperone\ we argued that these two effects are related.

Another effect seen in Figures~\ref{f.act-rot1} and \ref{f.act-rot-nofl} is
that the overall scatter in X-ray activity among rapid rotators is larger in
UCDs than in solar-type stars. For solar-type stars in the
\citet{Pizzolato2003} sample with $\prot < 1$~d, the standard deviation in
activity is $\sigma([\LxLb]) = 0.25$, while for UCDs in the same rotation
range, $\sigma([\LxLb]) = 0.85$. This effect is most pronounced when rotation
is parametrized in terms of \ro, which has less scatter in the pre-saturation
regime than \prot\ or $\rco/R_*$.

\subsection{Isolating the Role of Rotation}

It is important to understand whether the decrease in UCD X-ray activity is
caused by faster rotation or is due to trends in other stellar parameters such
as temperature, mass, or age. To this end we analyze the relationship between
X-ray activity and rotation in a subsample of UCDs in the narrow spectral type
range M6.5--M9.5. This subsample removes the influence of effective
temperature on activity, as \teff\ decreases by only \apx500~K between M6.5
and M9.5 \citep{Luhman2003}.

We analyze only nonflaring emission and remove known tight binaries from the
sample, consistent with the analysis of \citet{Pizzolato2003}. \citet{kgf+12}
showed that, in some UCD binaries, \vsi\ can differ significantly (by up to 30
\kms) between components, and it cannot be determined from available data
alone whether X-ray activity or rotational velocity are influenced by a
blending of the components.

\begin{figure}[hbt]
\plotonebig{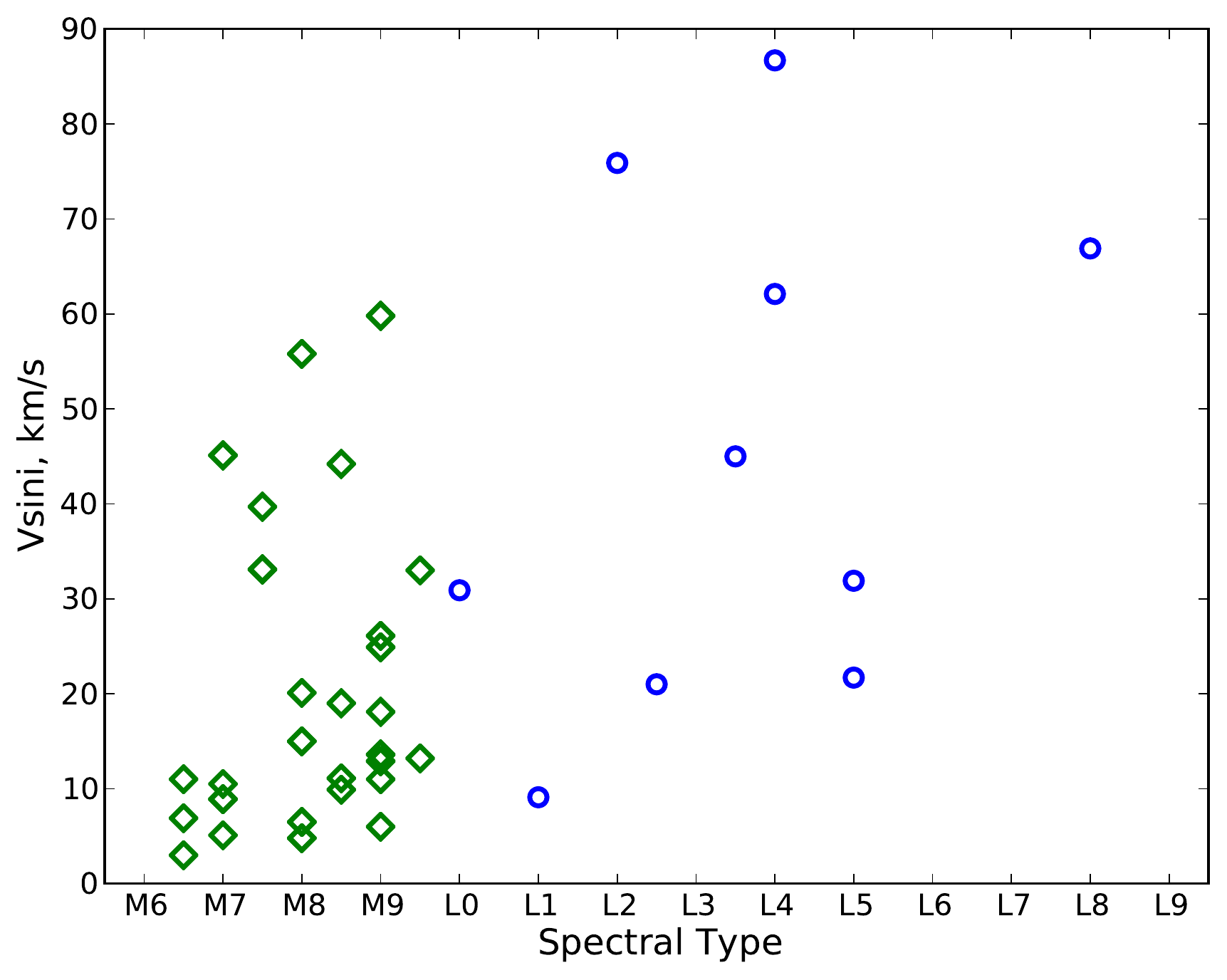}{0.95}
\caption{Projected rotational velocity in the UCD sample as a function of
  spectral type. Blue symbols represent L~dwarfs and green symbols represent
  M~dwarfs. L~dwarfs are systematically faster rotators. However no strong
  correlation with spectral type exists over the narrow spectral type regime
  M6.5--M9.5. Our analysis thus incorporates a subsample of UCDs in this
  range, in order to isolate the influence of spectral type (temperature) from
  purely rotational dependences.}
\label{f.vsini}
\end{figure}

In main-sequence low-mass stars, mass and rotation are correlated
\citep{ib08}. However, as shown in Figure~\ref{f.vsini}, \vsi\ is not strongly
correlated with spectral type within our subsample. Furthermore, the new
observations presented in \paperone\ contribute to the subsample at a wide
range of \vsi, so our dataset is capable of isolating the role of
rotation despite its restriction to a narrow range of \teff.

We perform regressions to investigate correlations between the various
rotational parameters and X-ray activity. Several UCDs in our sample
(particularly L dwarfs) are not detected in X-rays, requiring the use of
survival analysis methods \citeeg{Feigelson2012}. Our regressions use a
maximum likelihood estimator (MLE) method based upon an accelerated
failure-time model (method \textsf{cenmle}) in the package NADA
\citep{Helsel2005} in the R language. One benefit of this method is that it
allows for automatic calculation of the likelihood ratio of the parametrized
model to a model without dependence on rotation, yielding a statistical
$p$-value that indicates the strength of the correlation. To estimate the
uncertainty in the fit parameters, we performed a bootstrap analysis, running
$10,000$ iterations on the fit, sampling with replacements and calculating the
MLE fit parameters for each sample. The distributions of fit parameters for
all regressions are approximately normal, so the standard deviation of the
distribution is adopted as the uncertainty of each fit parameter.

\begin{deluxetable}{llccc}
\tablecolumns{5}
\tabletypesize{\small}
\tablecaption{X-ray Activity-Rotation Regression Parameters}
\tablehead{\colhead{$X$} & \colhead{$Y$} & \colhead{$a$} & \colhead{$b$} & \colhead{Prob.}}
\cutinhead{Full UCD sample}
\vsi\ (\kms)  & \Lx   & $-1.3 \pm 0.5$ & $27.0 \pm 0.6$ & $p<0.02$ \\
\vsi\ (\kms)  & \LxLb & $-0.9 \pm 0.4$ & $-3.5 \pm 0.5$ & $p<0.04$ \\
\prot\ (days) & \LxLb &  $1.0 \pm 0.4$ & $-4.1 \pm 0.3$ & $p<0.02$ \\
\ro           & \LxLb &  $1.0 \pm 0.4$ & $-2.2 \pm 0.9$ & $p<0.02$ \\
\cutinhead{Subsample: M6.5--M9.5 dwarfs, no binaries}
\vsi\ (\kms)  & \Lx   & $-1.3 \pm 0.6$ & $27.2 \pm 0.6$ & $p<0.03$ \\
\vsi\ (\kms)  & \LxLb & $-1.0 \pm 0.6$ & $-3.4 \pm 0.6$ & $p<0.07$ \\
\prot\ (days) & \LxLb &  $1.0 \pm 0.6$ & $-4.1 \pm 0.3$ & $p<0.06$ \\
\ro           & \LxLb &  $1.0 \pm 0.6$ & $-2.3 \pm 1.2$ & $p<0.06$
\enddata

\tablecomments{Regression parameters of $[Y] = a[X] + b$, derived using the
  MLE method described in \S\ref{s.trends}. While significant scatter exists
  in the data, the likelihood analysis allows us to dismiss no correlation
  with rotation with reasonable confidence. P-values are calculated from
  $\chi^2$ values derived by the regression algorithm. Interestingly, the fit
  for the entire UCD sample is consistent with the subsample of M6.5--M9.5.
  This indicates that, while later spectral types are systematically faster
  rotators, spectral type variations alone are not entirely responsible for
  the observed correlation; rotation has a significant influence on X-ray
  emission.}
\label{tab:regress}
\end{deluxetable}

\begin{figure}[hbt]
\plotonebig{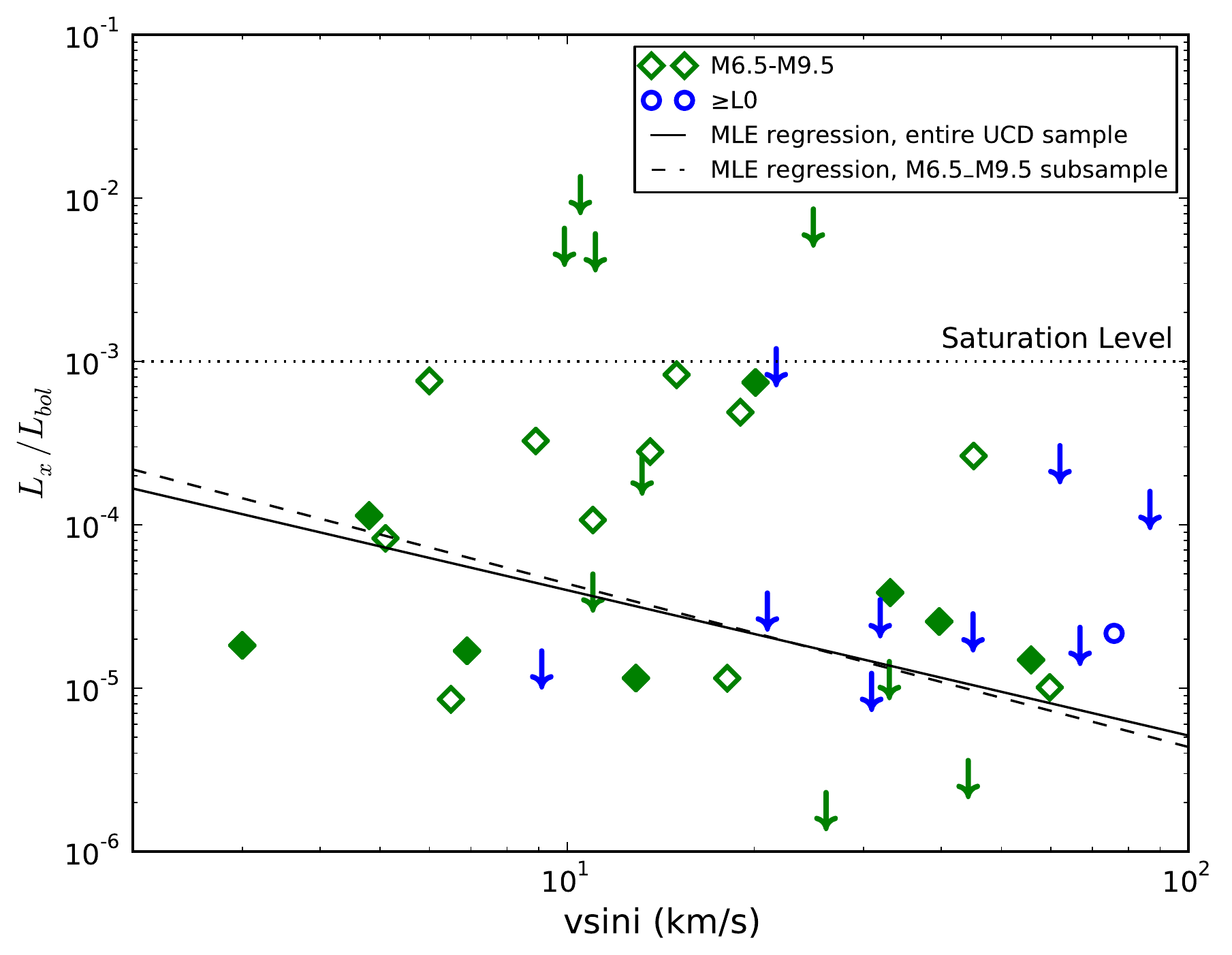}{0.95}
\caption{The UCD activity-rotation relation in terms of \vsi. X-ray activity
  is converted to the 0.2--2.0 keV band (see \paperone). Symbols and colors
  are the same as in Figure~\ref{f.act-rot1}. The \emph{solid line} shows the
  MLE regression ($p < 0.07$) for the M6.5--M9.5 subsample that isolates the
  effects of rotation from the those of the large mass and temperature changes
  that occur across the full UCD sample. The \emph{dashed line} shows the
  regression across the full sample ($p < 0.04$), which is statistically
  indistinguishable.}
\label{f.act-rot-M}
\end{figure}

The results of the survival analysis regressions are shown in
Table~\ref{tab:regress}. Although there is significant scatter in the data,
each regression finds that X-ray activity decreases with faster rotation in
UCDs. The $p$-values of the regressions range between $p < 0.02$ and $p <
0.07$. Figure~\ref{f.act-rot-M} shows the regressions of $\LxLb$ against
\vsi\ derived for both the entire UCD sample and the subsample of M6.5--M9.5
dwarfs. In this and all other cases, the regressions in both the full sample
and the rotation-isolating subsample are statistically indistinguishable,
which is consistent with a scenario in which there is no correlation between
\teff\ and \LxLb\ \emph{within the UCD regime}. However, it is clear that
\teff\ strongly affects \LxLb\ in the sense that UCDs behave very differently
than \emph{earlier-type objects with comparable levels of rotation}: the mean
level of their X-ray activity is significantly lower ($\LxLb \approx
10^{-4.5}$ versus $10^{-3}$), there is significantly more scatter in
\LxLb\ within the population ($\sigma([\LxLb]) \apx 0.9$ versus $0.3$), and
there is an anticorrelation between \LxLb\ and rotation as opposed to a null
correlation.

\section{Possible Causes of Decreased X-ray Emission in UCDs}
\label{s.disc}

Our analysis in \S\ref{s.trends} found evidence for an anticorrelation between
X-ray activity and rotation in UCDs, in both in the total UCD sample and in a
subsample of M6.5--M9.5 stars, where rotation is not strongly correlated with
spectral type and temperature changes by less than \apx20\%. We believe this
to be the strongest evidence to date of a breakdown in the standard X-ray
activity-rotation relation in UCDs. We now consider possible explanations for
this reduction in X-ray activity in UCDs, which we group into two broad
categories: explanations that argue for a ``supersaturation'' mechanism with a
causal relation between rapid rotation and reduced activity levels; and
``anticorrelations'' in which an additional parameter independently affects
both rotation and activity, leading to the observed trend.

\subsection{Supersaturation Mechanisms}
\label{s.supersat}

Several explanations for supersaturation of magnetic activity in ultra-fast
rotators have been proposed previously, motivated by the mild drop in
\Lx\ (factors of \apx2--3) seen in the most rapidly-rotating solar-type (G/K)
stars. In these stars the rotationally-driven $\alpha\Omega$ dynamo is
believed to generate magnetic activity, and it has been suggested that
supersaturation may be due to negative feedback in the dynamo at extremely
fast rotation rates \citep{krk94}. The $\alpha\Omega$ dynamo is not believed
to operate in fully-convective UCDs, and a turbulent dynamo is expected to
have only a mild dependence on rotation \citep{Durney1993}. A
rotationally-dependent $\alpha^2$ dynamo, on the other hand, might show
supersaturation effects at high rotation, possibly explaining the decrease in
coronal emission found here in UCDs. However, radio emission in some UCDs does
not saturate or supersaturate with fast rotation \citep{bbf+10,mbr12}, as
coronal and chromospheric activity appear to. Instead, radio activity in stars
as late as L4 increases with rotation to several orders of magnitude above the
saturation level of solar-type stars, implying continued, robust operation of
a magnetic dynamo.

Another proposed cause of supersaturation is centrifugal stripping of the
coronal envelope \citep{Jardine1999,James2000,Jardine2004}. This model
suggests that centrifugal forces in the outer coronae of rapidly rotating
stars lead to an increase in coronal pressure and density, increasing the
X-ray emissions coming from a given coronal volume. The volume of the emitting
coronal regions will, however, be limited by the corotation radius.
\citet{Jardine1999} argue that, in the moderate-rotation regime, the two
effects will cancel, leading to X-ray saturation. However, at sufficiently
rapid rotation periods the coronal stripping will reduce the overall X-ray
emission, causing a supersaturation effect.

The point at which centrifugal stripping would result in
supersaturation depends on the characteristic height of magnetic field
loops in the corona.  \citet{Wright2011} claim to find evidence for
supersaturation (activity decrease from $\LxLb \approx 10^{-3}$ to
\apx$10^{-3.5}$) in solar-type stars only at values of $\rco / R_*
\approx 2$--3. Yet, as shown in Figure~\ref{f.act-rot1}, UCDs with
suppressed X-ray emission in our sample have values of $\rco / R_*$
even higher than 10, in the same range as solar-type stars which show
saturated emission. If centrifugal stripping is responsible for
supersaturation in UCDs, the relative sizes of their coronal loops
must be $\gtrsim$5 times larger than even early M~dwarfs. Furthermore,
the magnitude of the observed X-ray activity decrease (nearly two
orders of magnitude) is much larger than that observed in solar-type
stars. We therefore do not find it plausible that centrifugal
stripping is the chief mechanism behind the UCDs supersaturation
effect.

\subsection{Sources of Anticorrelation}

The anticorrelation between activity and rotation does not imply a causal
connection between the two parameters. Although our analysis indicates a
general trend towards lower activity with faster rotation, the existence of
pairs of stars with similar spectral types and rotation velocities yet vastly
different X-ray activity levels challenges a causal supersaturation
interpretation. For example, the M9 dwarfs \objl{lhs2065} and
\objl{2massj10481463-3956062} ($\vsi = 13.5$ and $18$~\kms, respectively)
likely have similar masses, temperatures, and rotational velocities, yet
differ by a factor of \apx30 in X-ray activity ($[\LxLb] = -3.55$ and
$-4.95$). Indeed, the overall scatter in X-ray activity among rapid rotators
is much larger in UCDs than in solar-type stars. We speculate that an
additional parameter, weakly correlated with both rotation and activity, is
responsible for this large scatter and leads to the observed anticorrelation.

One possible parameter is the effective temperature \teff. The overall
decrease in the X-ray emission of UCDs has been attributed to
declining values of \teff, via increasingly neutral photospheres that
couple to the magnetic field progressively less effectively
\citep{Mohanty2002}. However, it is difficult to understand how such a
mechanism alone could explain the rotational dependence of X-ray
activity found in the current sample of active UCDs, which is observed
even in the M6.5--M9.5 subsample that isolates a population of similar
temperatures and masses. Purely temperature-dependent arguments
additionally cannot explain the widely varying X-ray activity among
stars with similar spectral types. We therefore do not consider \teff
as the relevant parameter.

Recent spectropolarimetry results suggest that the topology of stellar
magnetic fields changes significantly in the fully-convective
regime. These results come from the use of Zeeman-Doppler imaging
(ZDI) techniques \citep{thezdi}, which measure Zeeman splitting in the
Stokes~V line profiles and are thus sensitive to the net magnetic
flux over a resolution element (denoted $B_V$), rather than the total
field strength ($B_I$). \citet{Morin2008b} used ZDI to map the
large-scale magnetic fields of six fully-convective M3--M4.5 stars,
finding in all cases a strong mean field, \mbv, in a low-multipolar
configuration. Continuing this analysis into the spectral type range
M5--M8 (including \objl{g208-45}, \objl{vb8}, and \objl{vb10} from our
sample), \citet{mdp+10} found that late-M stars exhibit either
strong large-scale magnetic fields (similar to mid-M stars) or
fields with weak large-scale components, dominated by small-scale
fields (similar to solar-type stars). A key result was that both
topologies were observed in stars with similar masses and rotation
periods. Numerical modeling suggests that two separate dynamo regimes,
producing magnetic fields with differing topologies, could be mutually
stable in low-mass stars, leading to the observed bimodality in
magnetic field structure \citep{Morin2011, Gastine2013}. \citet{mbr12}
argued for the existence of a similar bimodality based on radio
observations.

\begin{figure}[hbt]
\plotonebig{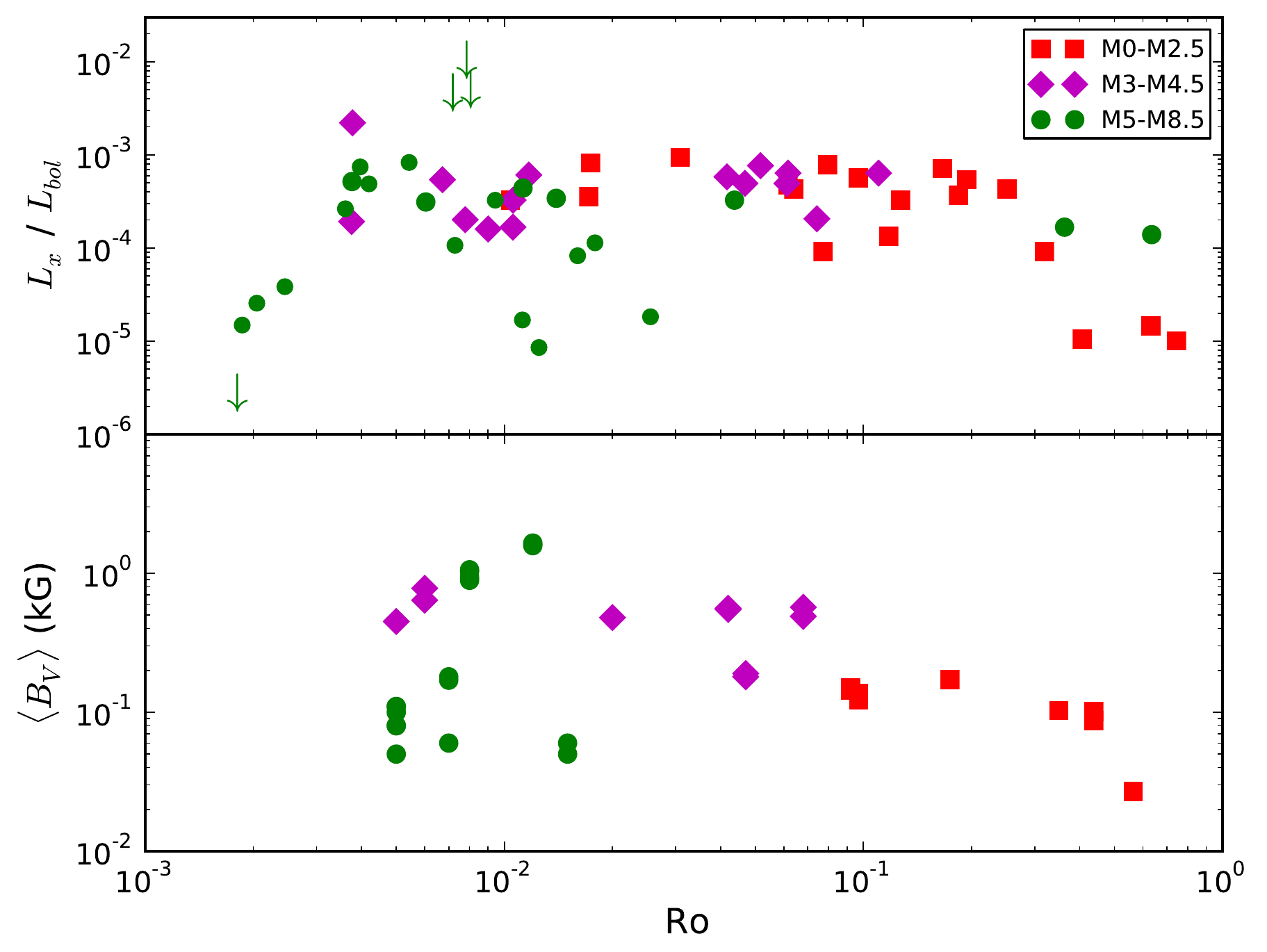}{0.95}
\caption{Similarity in the Rossby-number dependence of \LxLb\ and \mbv.
  \textit{Upper panel:} reproduction of the nonflaring data points from the
  middle panel of Figure~\ref{f.act-rot1}, with a slightly altered color
  coding. \textit{Lower panel}: reproduction of the ZDI data reported by
  \citet{dmp+08} and \citet{Morin2008b,mdp+10}. Although only three sources
  overlap between the samples (see text), both panels show similar trends:
  negative correlation in early-M~dwarfs (\textit{red}), flat relationship in
  mid-M~dwarfs (\textit{magenta}), and scattered, positive correlation in
  late-M~dwarfs (\textit{green}).}
\label{f.rxbv}
\end{figure}

Figure~\ref{f.rxbv} shows that the large-scale magnetic field strengths for
the objects studied in \citet{mdp+10} correlate with rotation and spectral
type in a remarkably similar fashion to the observed X-ray
activity/rotation/spectral-type relations in M~dwarfs. Mid-M~dwarfs (mostly
rapid rotators with $\ro < 0.1$) have ``saturated'' large-scale field
strengths (\apx600~G) independent of rotation. Late-M~dwarfs ($\ro \lsim
10^{-2}$) show large scatter in their large-scale strengths, with some
occupying the saturated branch and others falling as much as an order of
magnitude below the saturated level. This is strongly reminiscent of the the
behavior seen in the measurements of \LxLb\ presented in this work.

In \paperone, we suggested that the topology of the magnetic field affects
radio and X-ray luminosity ratios, with large-scale fields being associated
with relatively high X-ray activity levels. The trends in X-ray activity and
rotation that we describe here are consistent with this proposal. In this
scenario, the large scatter in X-ray activity as a function of rotation is due
to the varying strength of the large-scale stellar field components.
Differences in the X-ray emission from stars with similar rotational
velocities and spectral types (such as \objl{lhs2065} and
\objl{2massj10481463-3956062}) may trace the relative topology of their
magnetic fields.

The physical basis for such a connection is not fully obvious, because
well-ordered, large-scale fields might be expected to be less likely to
generate the reconnection events that are thought to ultimately power coronal
X-ray emission. We suggested in \paperone\ that the objects with small-scale
fields may generate many small reconnection events that are insufficiently
energetic to produce X-ray emission in the standard chromospheric evaporation
model. We also emphasize that ZDI measurements are sensitive to only some
aspects of the full complexity of stellar magnetic fields. In particular,
unresolved oppositely-directed field components --- such as the footpoints of
compact coronal loops --- cancel out in ZDI data, and observations of FeH
molecular lines suggest that ZDI measurements are sensitive to $\lesssim$15\%
of the total magnetic field \citep{rb09b}. A more detailed physical treatment
must handle such ``blind spots'' carefully.

The anticorrelation between rotation and X-ray activity will be
enhanced if stars with large-scale fields also exhibit generally
slower rotational velocities. Although \citet{Gastine2013} found that
large-and-small scale topologies can be generated in the same rotation
regime, stellar rotational evolution may depend on the large-scale
magnetic topologies. Spin-down processes may be affected by the
increased moment of inertia from an extended corona, and magnetic
braking is believed to be more significant in large-scale fields than
in tangled, high-multipolar fields \citep{Reiners2012a}. We speculate
that stars with large-scale, dipolar fields may form with similar
rotation values as low-activity, multipolar stars but spin down more
quickly, reinforcing the observed anticorrelation between activity and
rotation.

Little overlap exists between UCDs observed in X-ray and those with measured
magnetic topologies. \citet{mdp+10} find that \objl{gj1245b} and \objl{vb10}
($\LxLb = -4.3$ and $-5.1$ in quiescence) show no evidence of strong
large-scale components, consistent with our interpretation. \objl{vb8} is also
found to be dominated by small-scale fields, although it shows higher
quiescent activity ($\LxLb = -3.5$). \citet{mbi+11} suggest that observed
periodicities in radio emission in \objl{n33370} ($\LxLb = -3.6$ in
quiescence) suggest the existence of significant large-scale field components,
although this has not been confirmed with ZDI observations. Further study is
required to examine the connection between magnetic topology and coronal
activity measurements in UCDs, although this is complicated by the difficulty
of obtaining ZDI measurements of rapidly-rotating stars.

\section{Summary and Conclusions}
\label{s.conc}

We present the most complete sample to date of ultracool dwarfs
($\gtrsim$M7) with measured rotational velocities and X-ray luminosities.
Included in this sample are new measurements from \paperone. Combining these
with data available from the literature, our sample contains 38 objects later
than M6. All of these UCDs are fast rotators ($\prot\lsim1$~day) yet the
majority show X-ray luminosity at least an order of magnitude lower than
predicted from their bolometric luminosities and the established saturated
activity-rotation relation for early-M dwarfs.

We find evidence for an anticorrelation between X-ray activity and rotation in
the UCD sample that is much more extreme than the supersaturation effects
observed in rapidly-rotating solar-type stars, suggesting that UCDs do not
merely represent the rapidly-rotating end of the same activity-rotation
relation found in solar-type stars. This conclusion is reinforced by the much
larger scatter in X-ray activity levels found in UCDs compared to earlier-type
stars.

As radio activity was found by \citet{mbr12} to increase with faster rotation
even in the UCD regime, it is unlikely that the decrease in X-ray activity is
due to a decrease in the effectiveness of the magnetic dynamo. Centrifugal
stripping alone is also unlikely to be responsible, as the supersaturated UCDs
would require extremely large coronae for centrifugal effects to be
significant. The overall scatter in quiescent activity between stars with
similar spectral types and rotational velocities suggests that a separate
parameter, correlated weakly with both rotation and X-ray activity, is
required to match the observed trends.

We suggest that the magnetic field topology represents such a
parameter, as large-scale field strengths in convective stars
correlate with rotation and spectral type in a remarkably similar
fashion to X-ray activity. Ultracool Dwarfs with large-scale fields
may spin down more efficiently than those with weaker, tangled fields,
enhancing the observed anticorrelation.

This framework can be related to the observed correlations between radio and
X-ray emission in UCDs. As discussed in \paperone, there is evidence that UCDs
can be divided into two groups: objects that are radio-bright and X-ray-faint,
and those that are radio-faint but X-ray-bright. In our proposed scenario, the
former would harbor weak tangled fields, while the latter would harbor
stronger, dipolar fields. The bimodality observed in the latest-type objects
implies that rapid rotators can be radio-faint and X-ray bright, but that slow
rotators ($\ro \gtrsim 0.1$) are not expected to reach the extremely high
radio/X-ray flux ratios (\apx$10^4$ above expectations) seen in some UCDs.

Further observations are required to increase the number of UCDs with known
magnetic topologies and X-ray fluxes, but we speculate that X-ray activity
should be correlated with the large-scale fraction of the magnetic fields.
Further ZDI measurements (performed in conjunction with X-ray observations)
can help illuminate the effect of large-scale topology on coronal emissions.
Obtaining an improved understanding of the radio properties of the fastest
rotators should also be a high priority. The newly upgraded Karl G. Jansky
Very Large Array, with its sensitivity increase of nearly an order of
magnitude compared to the original system, offers an excellent opportunity to
improve on previous work. In addition, theoretical models should continue to
push into extremely-rapid rotation regimes, to provide accurate explanations
for the trends being found observationally.

\acknowledgments B.~A.~C.~thanks Jonathan McDowell and Marie Machacek for
their advice and comments on drafts of this work. This work is supported in
part by the National Science Foundation REU and Department of Defense ASSURE
programs under NSF Grant no. 1262851 and by the Smithsonian Institution.
E.~B. and P.~K.~G.~W. acknowledge support for this work from the National
Science Foundation through Grant AST-1008361, and from the National
Aeronautics and Space Administration through Chandra Award Number GO2-13007A
issued by the Chandra X-ray Observatory Center, which is operated by the
Smithsonian Astrophysical Observatory for and on behalf of the National
Aeronautics Space Administration under contract NAS8-03060. This research has
made use of the SIMBAD database, operated at CDS, Strasbourg, France, and
NASA's Astrophysics Data System.

\bibliography{xrot,../bib/pkgw,../lxlr/extra}

\end{document}